# The brain as a trigger system

**Maria Michela Del Viva**

*NEUROFARBA Dipartimento di Neuroscienze, Psicologia, Area del Farmaco e Salute del Bambino Sezione di Psicologia, Università di Firenze*
*Via di San Salvi, 12 Complesso di San Salvi, Padiglione 26, 26 - 50135 Firenze, Italy.*
*E-mail:* `maria.delviva@unifi.it`

**Giovanni Punzi**

*Dipartimento di Fisica "E. Fermi" Università di Pisa, via Buonarroti, 2 56127 Pisa, Italy.*
*E-mail:* `giovanni.punzi@pi.infn.it`

There are significant analogies between the issues related to real-time event selection in HEP, and the issues faced by the human visual system. In fact, the visual system needs to extract rapidly the most important elements of the external world from a large flux of information, for survival purposes. A rapid and reliable detection of visual stimuli is essential for triggering autonomic responses to emotive stimuli, for initiating adaptive behaviors and for orienting towards potentially interesting/ dangerous stimuli. The speed of visual processing can be as fast as 20 ms, about only 20 times the duration of the elementary information exchanges by the action potential.

The limitations to the brain capacity to process visual information, imposed by intrinsic energetic costs of neuronal activity, and ecological limits to the size of the skull, require a strong data reduction at an early stage, by creating a compact summary of relevant information, the so called "primal sketch", to be handled by further levels of processing. This is quite similar to the problem of experimental HEP of providing fast data reduction at a reasonable monetary cost, and with a practical device size.

As a result of a joint effort of HEP physicists and practicing vision scientists, we recently proposed that not only the problems are similar, but the solutions adopted in the two cases also have strong similarities, and their parallel study can actually shed light on each other.

Modeling the visual system as a trigger processor leads to a deeper understanding, and even very specific predictions of its functionality. Conversely, the insights gained from this new approach to vision, can lead to new ideas for enhancing the capabilities of artificial vision systems, and HEP trigger systems as well.

# 1. Introduction

This work started a few years ago from the intuition that advanced data acquisition systems in experimental High Energy Physics (HEP) deal with data processing problems similar to those faced by the human visual system [1]. In the studies that followed, we found out that the analogy is more than superficial, and we learned a number of interesting things through a multidisciplinary study of both systems [2]. The main theme underlying these studies is that the effort for optimization of artificial systems for data acquisition in HEP on one side, and the evolutionary pressure for the optimization of the visual system on the other side, has led to a common set of solutions to some data processing problems - a process that could be described as a "convergent evolution" of natural and man-made information-processing systems.

In this report, we describe these studies in detail as a case study of an interaction of two previously separated disciplined that sparked some interesting new ideas in both fields.

## 1.1 Functional similarities between data reduction in HEP experiments and early visual processing

HEP experiments in large hadron colliders like the LHC, or the Tevatron before it, generate large flows of data. Typical rates of particle collisions of several MHz, each producing O(Mb) of data, lead to data flows measured in Terabytes per second. The approach used to deal with such large data flows is to perform a real-time selection of a small fraction of interesting events (typical $<10^{-3}$), and only save that small fraction on permanent storage for later (*off-line*) analysis. This selection process requires to make critical decisions, often requiring a complex analysis of a significant fraction of the event data. To do such complex analysis in a short time, dedicated *trigger devices,* and special techniques, have been developed. In particular, extraction of meaningful information requires a strong data reduction process internal to the trigger.

The SVT is a successful example of a complex trigger device, used in reconstructing charged particles trajectories going through the CDF detector at the Tevatron collider in timescales of ~10μs [3]. Track reconstruction is based on a pattern-matching algorithm, implemented by using a custom VLSI device: the *Associative Memory* [4]. Associative memories were devised for this specific trigger but are still in use and under development for current tracking systems (FTK for ATLAS at LHC [5], and ongoing R&D for CMS). Patterns in the AM are pre-calculated sequences of coordinates in the detector layers, corresponding to plausible particle trajectories. Data coming from the detector are read out sequentially into the AM, and compared in parallel to the entire set of patterns stores in the AM. Only the data matching the elements in the pattern bank are retained and transmitted to the following stages of trigger processing for further analysis, for extraction of precise measurements of particle parameters, eventually leading to a decision on retaining or discarding the whole event. The huge data reduction operated by the SVT allowed CDF to produce important physics results on heavy quark decays, that were previously accessible only in experimental setups with much smaller data flows [3].

The visual system in mammals' brain also needs to extract relevant information from a large amount of input data in real time, for survival purposes. Fast and reliable detection of visual stimuli is essential for triggering autonomic responses to emotive stimuli, for initiating adaptive behaviors and for orienting towards potentially interesting/dangerous stimuli. The data reduction operated by the visual system is strong. At retinal level, information is processed at about 20 Gb/s by $10^8$ photoreceptors, while output information is provided at a rate between 0.8-4 Gb/s by about $10^6$ Ganglion cells, whose axons form the optic nerve [6]. A second, even tighter bottleneck, is thought to operate at early cortical level [7] and is estimated as having a final capacity of just 40 bits/second for humans [8]. In terms of processing speed, neuroimaging and psychophysical studies on human subjects have shown that the Primary visual area (V1) can do a first analysis of an image in times as short as 20-30 ms [9,10]. To compare the demands made on algorithms and processor architectures by these numbers in the visual and the HEP world, it is necessary to account for the relative speed of the processing hardware. The elementary time unit of the visual system is given by the typical switching time of neurons, which is of order 1 ms; this can be compared to current electronics switching times or order 1 ns, leading to a factor of order $10^6$ between the hardware speed of the two systems. Looking at the above numbers, it appears that the computational load of the visual system, normalized to the hardware speed, is quantitatively at least comparable to that of the LHC experiments in the high-luminosity conditions expected in the upcoming data taking periods. There are other functional similarities between the two systems, up to practical limitations imposed by intrinsic costs – some are displayed in Table 1.

| NATURAL VISION | HEP DAQ |
| --- | --- |
| Extensive early data reduction [11,12] | Limited Offline storage |
| A strong, lossy data-reduction [7] | Use strongly reduced information |
| Size of brain limited | Size of device limited by cost |
| Limits to energy consumption [13,14] | Limits to electrical power |
| Number of visual neurons and their discharge rate not sufficient to process all data [15, 6]. | Bandwidth and computing power at higher trigger levels cannot process the full rate. |

**Table 1** Computational limitations in early vision and data acquisition in HEP

It has long been proposed that the visual system solves the data-reduction problem by creating a compact summary of the image ("primal sketch"), based on few simple features, to be handled by further levels of processing [16].

Given the analogies with HEP triggers, where a similar problem has been successfully solved with a hardware-implemented pattern-matching algorithm, we explored the idea of deriving the primal sketch with the same method. This led to a detailed work [2] that we summarize here in brief.

**2. The problem of optimal data reduction in pattern matching**

Let us consider a generic Information Processing System that receives in input a large amount of data and is expected to provide in output a "summary" of the input information for another device to perform further processing. We postulate that the summary is based on recognizing a limited number of meaningful patterns of the input, dropping the remaining information (pattern matching). We also assume that the system is subject to some computational limitations: the matching is done with a fixed number of stored patterns (finite memory/storage) and the output bandwidth is fixed by limitations of the next stage.

The functionality of such abstract pattern-filtering model is completely defined by its reference set of patterns. While in HEP the nature of the desirable patterns is completely known (the particle tracks), for vision it is not obvious what information is actually used for the summary/sketch. One may however ask the question of what is the set of stored patterns that such system ought to use, to provide the best possible summary in output. One could expect that a critical system that has been subject to extensive evolutionary pressure should have reached a point very close to optimality. Clearly, one needs a precise definition of what should be optimized. We adopted the viewpoint that optimality consists in delivering the maximum amount of information to the following processing stages, within the assigned external constraints. This can be considered as a working hypothesis, to be reevaluated on the basis of the results it produces. It has the great advantage over more specific definition, of allowing to address the issue in general, without specific knowledge of the field, simply by looking for a maximum-entropy solution of the problem at hand. Having chosen a discrete representation, the output entropy can be expressed as $\sum_{i}^{N} -p_i \log(p_i)$, where N is the number of patterns and $p_i$ is the probability of occurrence of each pattern $i \in$ N. We can associate a "cost" to each pattern, defined as the larger of the "storage cost" 1/N and the "bandwidth cost" $p_i/W$, where W is the maximum allowed total rate of pattern acceptance, S $p_i$ < W. As a consequence, an entropy yield per unit cost is given for each pattern by:

$$f(p) = \frac{-p \, \log(p)}{\max(1/N, p/W)} \qquad (2.1)$$

The optimal performance of the filtering system is then attained by choosing the set of patterns such that $f(p_i) > c$, where $c$ is determined by the computational limitations.

**3. Application to real-time data acquisition in HEP**

The concepts outlined in the previous section can be applied to data-reduction in HEP experiments. We used a sample of simulated data with a Monte Carlo method, assuming for simplicity the earliest and simplest configuration for a detector structure that has been in actual use in particle physics experiments, with five planar measuring layers [3]. From the knowledge of the detector geometry and response characteristics, we computed a priori the set of all possible patterns corresponding to valid particle traversing the detector, as it is done in real AM applications [3]. In a second step, we generated the probability distribution of the frequency of all possible patterns, from a sample of 100,000 simulated events. We then proceeded to compare the probabilities of the valid patterns within the overall distribution with the prediction obtained from our own recipe outlined in the previous section (Fig. 1).

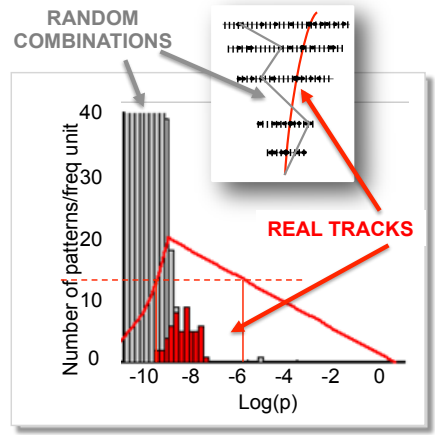

**Figure 1** Monte Carlo simulation of track reconstruction and pattern filtering in a HEP particle detector. **Inset**: Schematic representation of a sector of a five-layers tracking detector. Black dots represent measured positions (*hits*) produced by particles or random noise. Each layer is subdivided into intervals (*bins*). Every possible combination of bins defines a *pattern* (grey line), some of them compatible with real tracks (red line). The grey histogram is the probability distribution of the frequency of patterns produced by the sample of simulated events. The distribution of the sub-sample of patterns corresponding to valid particle trajectories is shown as a red histogram. The red curve is the function of eq. 2.1. The vertical red lines indicate the probability range selected by our model.

Comparison shows that the patterns corresponding to real particle tracks all fall within a limited range of intermediate probability within the overall distribution. This implies that they could have been identified and selected purely on the basis of a constrained maximum-entropy calculation (for an appropriate choice of parameters).

This is a first example where cross-contamination of concepts from different fields lead to new ideas. The effort of generalizing concepts so they can applied to different circumstances, led to a new view of particle tracks: they can be formally described as the piece of information that carries the maximum amount of information under the practical bandwidth and size limitations of the hardware. This interesting fact is not currently exploited in HEP applications, because one can usually rely on the a-priori knowledge of the relevant patterns - however, in principle it might find some applications in automatic adaptation to changes in detector alignment.

In any case, the success of this approach in identifying the right patterns in a HEP application is an important confirmation of its soundness, and had encouraged us to consider its application to the question of identification of salient features in natural vision, where the "right set of patterns" for the problem is not *a-priori* known.

**4. Extraction of optimal visual patterns from natural image statistics.**

In applying our model to vision, we have considered the simplest possible set of base patterns, defined as all possible configurations of 3*3 square pixel matrices in black-and-white images (1-bit depth). We then evaluated the probability distribution of the patterns in a set of natural images extracted from a public database [17]. Then we extracted the optimal set of patterns according to eq. 2.1. Interestingly, it turns out that the majority of the patterns selected by this algorithm (Fig. 2, green) can be classified as edges, bars, end-stops, or corner detectors of various orientations, closely resembling (within the limitations of a 3*3 grid) the spatial structure of receptive fields of neurons in primary visual areas [18]. Conversely, most of the patterns discarded by our selection have either an irregular structure resembling visual noise (Fig. 2, blue), or uniform luminance (Fig. 2, red), with lower resemblance to known visual features. Direct evidence that the human visual system actually assigns to the selected patterns a privileged role in its image-reconstruction process has been obtained by dedicated psychophysical measurements. Contrast sensitivities for the detection of single isolated patterns has been measured [2]. The results of a scan over the entire range of pattern probabilities found in natural images show that the contrast sensitivity of all subjects peaks within a limited probability range, in agreement with the predictions of our model (Fig. 2, plot).

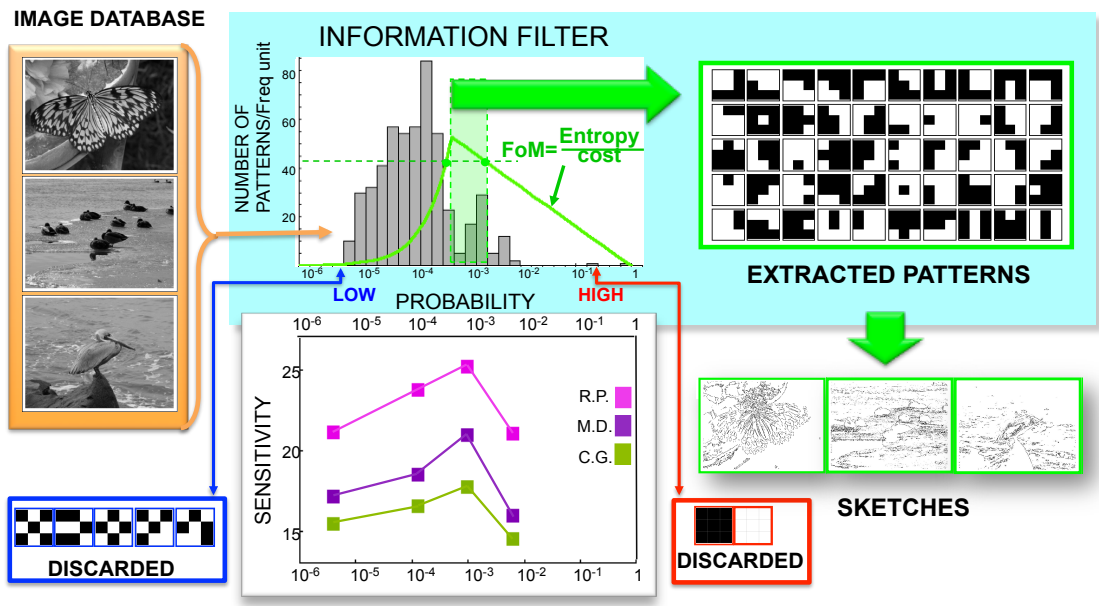

**Figure 2** Extraction of "optimal" visual patterns. The curve is the model selection function (eq. 2.1) for W=0.05 and N= 50 [2]. The bottom plot shows the average contrast sensitivity for detection of patterns as a function of their probability, on three human subjects (different colors).

Sketches extracted from our images by keeping only those patches of the binarized image matching one of the selected patterns (dropping all other parts) retain most of the salient features of originals, in spite of a substantial reduction of information, and elicit nearly the same response as complete images [2].

### 5. Discussion

In summary, the idea of performing a data-reduction via matching to a predefined set of patterns, born in experimental HEP, when carried over to the field of natural vision has led to the realization that the brain performs a closely similar algorithm in the early stages of the vision process, and to explain the observed shape of physiological receptive fields as the patterns that can be efficiently encoded within the available computational resources. While other models of features extraction [19, 20] or data reduction [21-23] have been formulated in vision science, none has yet been as predictive.

This achievement comes from having formulated the problem at a rather abstract level, focusing on the overall functionality of the systems, rather than procedural details. One important point that has been glossed over in the comparison, is the fact that the AM systems operated digitally on the data, while this is unlikely to be the case for the brain, where most data is represented as excitation levels of neurons, or frequency of their discharge, that can assume many different values, essentially akin to analog variables. For the purpose of the studies described above, this distinction is not important, as long as the final result of the processing is to decide on the presence or absence of a given pattern. This does not imply that a more detailed study, explicitly considering the analog nature of brain processing and other implementation details, is not worthwhile. A more explicit modeling of the process in terms of biologically-plausible situations can certainly lead to improved modeling and further understanding of the visual function. Considerations of continuous luminance levels, multiple scales, color, extension to time (motion), can, and are in fact being pursued [24, 25].

However, this can also reflect back on HEP experimental techniques. As a fallout of the previous discussion, it is natural to ask whether a more brain-like implementation of pattern-matching algorithms, based on continuous excitation levels rather than yes/no responses can be of some use in designing artificial data-processing devices. This kind of ideas have indeed being explored [26, 27], including some presentations in this same conference [28, 29], showing another way in which the interaction between the fields of natural vision and experimental HEP can generate interesting new developments.